\documentclass[11pt]{elsart}

\usepackage{enumerate}
\usepackage{amsmath}
\usepackage{amssymb}

\newcommand{\rr}{\ensuremath{\mathrm{I\!R}}}
\newcommand{\degr}{\ensuremath{\Delta}}
\newcommand{\boxi}{\ensuremath{\mathrm{box}}}

\newcommand{\tw}{\ensuremath{\mathrm{tw}}}

\newcommand {\bbox}{\rule{0.6em}{0.6em}}

\renewenvironment{pf}
{\begin{oldproof}}{\hfill\bbox\end{oldproof}}

\newcommand{\pp}[1]{\ensuremath{\mathbf{Pr}\!\left[#1\right]}}
\newcommand{\pe}[1]{\ensuremath{\mathbf{E}\!\left[#1\right]}}

\newcommand{\ignore}[1]{}
\newcommand{\com}[1]{}
\newcommand{\fignore}[1]{}

\newcommand{\surl}[1]{{\small\url{#1}}}

\begin{document}

\date{}
\begin{frontmatter}

\title  {Geometric representation of graphs in low dimension using axis parallel boxes  }

\author[sun]{L. Sunil Chandran}
and
\author[sun]{Mathew Francis}
\author[ns]{Naveen Sivadasan}

\address[sun]{Indian Institute of Science,
Dept. of Computer Science and Automation,
Bangalore  560012, India.  email: \emph{sunil,mathew@csa.iisc.ernet.in}}
\address[ns]{Advanced Technology Center, Tata Consultancy Services,
1, Software Units Layout, Madhapur, Hyderabad - 500081, India. email: \emph{s.naveen@atc.tcs.com}}


\maketitle
\pagestyle{empty}

\begin{abstract}

An axis-parallel  $k$--dimensional box is a Cartesian product
$R_1 \times R_2 \times \cdots \times R_k$ where $R_i$ (for $1 \le i \le k$)
is a closed interval of the form $[a_i, b_i]$ on the real line.
For a  graph $G$, its \emph{boxicity} $\boxi(G)$ is the minimum dimension $k$, such that
$G$ is representable as the intersection graph of
(axis--parallel) boxes in $k$--dimensional space. 
The concept of
boxicity finds applications in various areas such as ecology, operations
research etc.

A number of  NP-hard problems are either polynomial time solvable
or have much better approximation ratio on low boxicity graphs.
For example, the max-clique problem is polynomial time solvable 
on bounded boxicity graphs and the maximum independent set problem for
boxicity 2 graphs has
a $1+\frac{1}{c}\log n$ approximation ratio for any constant $c$.
In most cases, the first step
usually is computing a low dimensional box representation of the given graph.
Deciding whether the boxicity of a graph is at  most $2$ itself is NP-hard.

We give an efficient randomized algorithm to construct 
a box representation of any graph $G$ on $n$ vertices in $\lceil(\Delta + 2)
\ln n\rceil$
dimension, where $\Delta$ is the maximum degree of $G$.
This algorithm implies that $\boxi(G) \le \lceil(\Delta + 2) \ln n\rceil$
for any graph $G$.
Our bound is tight up to a factor of $\ln n$.
\ignore{
The only previously known general upper bound for boxicity
was given by Roberts, namely $\boxi(G) \le n/2$.
Our result gives an exponentially better upper bound
for bounded degree graphs.
}

We also show that our randomized algorithm can be derandomized
to get a polynomial time deterministic algorithm.

Though our general upper bound is in terms of maximum degree $\Delta$,
we show that for almost all graphs on $n$ vertices, their boxicity is
$O(d_{av}\ln n)$ where $d_{av}$ is the average degree.


\end{abstract}


\end{frontmatter}



\section{Introduction}

Let ${\mathcal{F}} = \{ S_x \subseteq U : x \in V \}$ be a family of 
subsets of a universe $U$, where $V$ is an index set. The intersection graph 
$\Lambda({\mathcal{F}})$ of ${\mathcal{F}}$ 
has $V$ as vertex set, and two distinct vertices
$x$ and $y$ are adjacent if and only if $S_x \cap S_y \ne \emptyset$.
Representations of graphs as the intersection graphs of various
geometrical objects is a well studied topic in graph theory. 
Probably the most well studied class of intersection graphs are
the \emph{interval graphs}, where each $S_x$ is a closed interval
on the real line.

A well known
concept in this area of graph theory is the \emph{boxicity}, which
was introduced by F. S. Roberts in 1969  \cite {Roberts}.
This concept generalizes the concept of interval graphs.
A $k$--dimensional box is a Cartesian product 
$R_1 \times R_2 \times \cdots \times R_k$ where $R_i$ (for $1 \le i \le k$)
is a closed interval of the form $[a_i, b_i]$ on the real line. 
For a  graph $G$, its \emph{boxicity} is the minimum dimension $k$, such that
$G$ is representable as the intersection graph of 
(axis--parallel) boxes in $k$--dimensional
space. We denote the boxicity of a graph $G$ by $\boxi(G)$. 
The graphs of boxicity $1$ are exactly the class of interval graphs.
The boxicity of a complete graph is $0$ by definition.


It was shown 
by Cozzens \cite {Coz}  that computing the boxicity of a graph is NP--hard. 
This
was later improved by Yannakakis \cite {Yan1}, 
and finally by  Kratochvil \cite {Kratochvil} 
who  showed that deciding  whether the
boxicity of a graph is at most 2 itself 
is NP--complete. 

In many algorithmic problems related to graphs, the availability of
certain convenient representations turns out to be extremely useful.
Probably, the most well-known and important examples are the tree decompositions
and path decompositions  \cite{Bodland3}. Many NP-hard problems are known 
to be
polynomial time solvable given a tree(path) decomposition of bounded width for 
the input graph. Similarly, the representation of
graphs as intersections of ``disks" or ``spheres" lies at the core
of solving problems related to frequency assignments in radio networks,
computing molecular conformations etc. For the  maximum independent set problem
which is hard to approximate  within a factor of $n^{(1/2) - \epsilon}$
for general graphs \cite{Has99}, a PTAS is known for disk graphs given the disk representation  
\cite{Erl01,Chan03}.
In a similar way,
the availability of box representation in low dimension
make some well known NP hard problems polynomial time solvable.
For example, the max-clique problem is 
polynomial time solvable in boxicity $k$ graphs  since there are only $O((2n)^k)$ maximal cliques
in such graphs.
It was shown in \cite{Has99} that the complexity of finding
the maximum independent set is hard to approximate within a factor $n^{(1/2) - \epsilon}$
for general graphs. In fact, \cite{Has99} gives the stronger inapproximability
result of $n^{1-\epsilon}$, for any $\epsilon>0$, under the assumption that
$NP\not=ZPP$. Though this problem is NP-hard  even for boxicity $2$ graphs,
it is approximable to a $\log n$ factor for boxicity $2$ graphs given a 
box representation \cite{Agarwal98,Berman2001}.


Thus, it is interesting to design efficient algorithms to 
represent small boxicity graphs in  low dimensions. 
\ignore{
To the best of our knowledge, the only known 
strategy  till date for computing a box representation for general graphs
is by Roberts \cite{Roberts}, but it guarantees only a box representation in
$n/2$ dimension for any graph $G$
on $n$ vertices and $m$ edges. In this paper, we give a
 randomized algorithm that guarantees an exponentially
 better bound ($O(\ln n)$ instead of $n/2$) for the dimension in case of bounded degree graphs.
 }
Roberts \cite{Roberts} had given a general upper bound of $n/2$ for the boxicity
of any graph on $n$ vertices. In this paper, we show an upper bound of
$\lceil(\Delta+2)\ln n\rceil$ for the boxicity
for any graph $G$ on $n$ vertices and having maximum degree $\Delta$
by giving a randomized algorithm that
\ignore{Our approach }yields a box representation for $G$ in 
$\lceil(\Delta +2 ) \ln n\rceil$ dimension in $O(\Delta n^2 \ln^2 n)$  time
with high probability. 
We also derandomize our randomized algorithm
and obtain a deterministic polynomial time algorithm to do the same.
Very recently, the authors had shown that for
any graph $G$ with maximum degree $\Delta$, $\boxi(G)\leq 2\Delta^2$ \cite{CFNMaxdeg}.
However, it may be noted that the result in this paper yields better bounds
on boxicity for graphs where $\Delta\geq\ln n$.

In a recent manuscript \cite{CN05} the authors showed that for any graph $G$,
$\boxi(G) \le \tw(G) + 2$,
where $\tw(G)$ is the treewidth of $G$. This result implies that
the class of `low boxicity'
graphs properly contains the class of `low treewidth graphs'. 
It is well known that almost all graphs on $n$ vertices and $m = c n$ edges
(for a sufficiently large constant $c$) have $\Omega(n)$ treewidth \cite{Klok2}. 
In this paper we show that 
almost all graphs on $n$ vertices and $m$ edges 
have boxicity $O(d_{av}\ln n)$ where $d_{av}=2m/n$.
An implication of this  result  is that
for almost all graphs on $m= cn$ edges, there is an exponential gap
between their boxicity and treewidth.
Hence it is interesting to reconsider those NP-hard problems that
are polynomial time solvable in bounded treewidth graphs
and see whether they are also polynomial time solvable for
bounded boxicity graphs.

Researchers have also tried to  bound the boxicity of 
graph classes  with special structure. 
Scheinerman \cite {Scheiner} 
showed that the boxicity of outer planar graphs is at most $2$.
Thomassen \cite {Thoma1} proved that 
the boxicity of planar graphs is
bounded above by $3$. Upper bounds for the boxicity of
many other graph classes such as chordal graphs, AT-free graphs, permutation graphs etc.
were shown in \cite{CN05} by relating the boxicity of a graph with its treewidth.
Researchers have also tried to generalize or extend  the
concept of boxicity in various ways. The poset boxicity \cite {TroWest}, 
the rectangle number \cite {ChangWest}, grid dimension \cite {Bellantoni},
circular dimension \cite {Feinberg,Shearer}  and the boxicity 
of digraphs \cite {ChangWest1} are some  examples. 

\subsection{Our Results}
We summarize below the results of this paper.

\begin{enumerate}

\item
We show that for any graph $G$ on $n$ vertices with maximum degree $\Delta$,
 $\boxi(G) \le \lceil(\Delta + 2) \ln n\rceil$.
This bound is tight up to a factor of $2 \ln n$.

\item
In fact, we show a randomized algorithm to construct a box representation
of $G$ in $\lceil(\Delta + 2) \ln n\rceil$ dimension, that runs in $O(\Delta
n^2 \ln^2 n )$ time with high probability.

\item
Next we show a polynomial time deterministic algorithm to construct
a box representation in $\lceil (\Delta + 2) \ln n \rceil$ dimension 
by derandomizing the above randomized algorithm.

\item
Though the general upper bound that we show is in terms of the maximum degree
$\Delta$, we also investigate the relation between boxicity and average degree.
We show that for almost all graphs on $n$ vertices and $m$ edges,
the boxicity is $O(d_{av} \ln n)$, where $d_{av} = 2m/n$ is the average degree.

\item
We also derive an upper bound for boxicity in terms of $m$ and $n$.
We show that for any connected graph $G$, $\boxi(G) \le 5 \sqrt{m \ln n}$,
which is tight up to a factor of $b \sqrt{\ln n}$ for a constant $b$.

\end{enumerate}

\subsection {Definitions and Notations}
Let $G$ be an undirected simple graph on $n$ vertices.
The vertex set of $G$ is denoted as $V(G) = \{1, \cdots, n\}$ (or $V$ in short).
Let $E(G)$ denote the edge set of $G$.
We denote by $\overline{G}$, the
complement of $G$. 
We say that the edge $e$ is missing in $G$, if $e \in E(\overline{ G})$.
A graph $G'$ is said to be a super graph of $G$ where $V(G) = V(G')$,
if $E(G) \subseteq E(G')$.
For a vertex $u \in V$, let $N(u)$ denote the set of neighbors of
$u$ in $G$ i.e., $N(u) = \{v \in V \,|\, u \not= v \mbox{~and~} (u,v) \in E(G) \}$.
Let $d(u)$ denote the degree of $u$ in $G$, i.e. $d(u) = |N(u)|$ .
Let $\Delta$ denote the maximum degree  of $G$.

\begin{defn}[Projection]
Let $\pi$ be a permutation of the set $\{1,\ldots,n\}$.
Let $X \subseteq \{1, \ldots, n\}$. The projection of $\pi$
onto $X$ denoted as $\pi_X$ is defined as follows.
Let $X = \{u_1, \ldots, u_r\}$ such that $\pi(u_1) < \pi(u_2) < ... < \pi(u_r)$.
Then $\pi_X(u_1) = 1, \pi_X(u_2) = 2, \ldots, \pi_X(u_r) = r$.
\end{defn}

\ignore{

	\begin{defn}
	Let $\pi$ be a permutation of the vertex set $\{1,\cdots,n\}$.
	Consider any two vertices $u$ and $v$ such that $(u, v) \in E(\overline{G})$.
	Then $\pi$ is said to be `good' for $(u, v)$ if the following is true.
	Let $X = \{u\} \cup N(u) \cup \{v\} \cup N(v)$ and let $\pi_X$ be the 
	projection of $\pi$ onto $X$. Let $i_u = \min_{w \in \{u\} \cup N(u)} \pi_X(w)$
	and let $i_v = \min_{w \in \{v\} \cup N(v)} \pi_X(w)$.
	Then $\pi$ is `good' for $(u, v)$ if and only if $r(v) < i_u$ or $r(u) < i_v$.
	Otherwise $\pi$ is bad for $(u, v)$.
	Let $G'$ be the interval super graph of $G$ defined by $\pi$. That is
	$G' = {\mathcal M} (G, \pi)$. 
	It is easy to see that if $\pi$ is good for $(u,v)$ then $(u, v) \in E(\overline{G'})$.
	Also, if $\pi$ is bad for $(u,v)$ then $(u, v) \notin E(\overline{G'})$.
	
	\end{defn}

}

\begin{defn}[Interval Representation]
An interval graph can be represented as the intersection graph
of closed intervals on the real line $\rr$. Such a representation is called
an interval representation. Equivalently and in fact more precisely, an interval representation
of an interval graph $G$ can be defined by two functions $l : V \rightarrow \rr$
and $r: V \rightarrow \rr$. The interval corresponding to a vertex $v$
denoted as $I(v)$ is given by $[l(v), r(v)]$, where $l(v)$ and $r(v)$
are the left and right end points of the interval corresponding to $v$.

\end{defn}

\begin {defn} \label{mmap}
Given a graph $G$ and a permutation $\pi$ of its vertices $\{1, \ldots, n\}$, we 
define a map ${\mathcal M}(G,\pi)$, which associates to the pair $(G, \pi)$, an
interval supergraph $G'$ of $G$, as follows:
consider any vertex $u \in V(G)$. Let $n_u \in N(u) \cup \{u\}$ be the vertex such
that  $\pi(n_u) =  \min_{w \in N(u) \cup \{u\}} \pi(w)$. Then
associate the interval $[\pi(n_u),\pi(u)]$ to the vertex $u$. Let 
$G'$ be the resulting interval graph. It is easy to verify that 
$G'$ is a super graph of $G$. We define ${\mathcal M}(G,\pi) = G'$. 
\end {defn}

\subsection{Box Representation and  Interval Graph Representation}
\label{sec:boxi-inter}

Let $G$ be a graph and let $I_1, \ldots, I_k$ be $k$
interval graphs such that each $I_j$ is defined on
the same set of vertices $V$. That is, $V(I_j) = V(G)$ for $j=1,\ldots,k$.
If $$E(G) = E(I_1) \cap \cdots \cap E(I_k),$$ then we say that
$I_1, \ldots, I_k$ is an \emph{interval graph representation} of $G$.
The following equivalence is well-known.

\begin{thm}[Roberts \cite{Roberts}]\label{thm:robertsx}
Let $G$ be a simple, undirected, non-complete graph.
Then, the minimum $k$ such that there exists an interval graph representation
of $G$ using $k$ interval graphs $I_1, \ldots, I_k$  is the same
as $\boxi(G)$.
\end{thm}

Recall that a $k$--dimensional box representation of $G$ is a mapping of each vertex $u \in V$
to $R_1(u) \times \cdots \times R_k(u)$, where each $R_i(u)$ is a closed interval of the
form $[\ell_i(u), r_i(u)]$ on the real line. It is straightforward to see that
an interval graph representation of $G$ using $k$ interval graphs $I_1, \ldots, I_k$,
is equivalent to a $k$--dimensional box representation in the following sense.
Let $R_i(u) = [\ell_i(u), r_i(u)]$ denote the closed interval corresponding to vertex $u$
in an interval representation of $I_i$.
Then the $k$--dimensional box corresponding  to $u$  is simply
$R_1(u) \times \cdots \times R_k(u)$. Conversely, given a $k$--dimensional box representation
of $G$, the set of intervals $\{R_i(u) : u \in V\}$ forms the $i$th interval graph $I_i$
in the corresponding interval graph representation.

When we say that a box representation in $t$ dimension is output by an algorithm,
the algorithm actually outputs the interval graph representation: that is,
the interval representation of the constituent interval graphs.

\section{The randomized construction}

Consider the following randomized procedure \textbf{RAND} 
which outputs an interval super graph of $G$.
Let $\Delta$ be the maximum degree of $G$.

\hfill

\noindent\textbf{RAND}


Input: $G$.

Output: $G'$ which is an interval super graph of $G$.

\noindent\textbf{begin}

    step1. Generate  a  permutation $\pi$ of $\{1,\ldots,n\}$ uniformly at random.

    step2. Return $G' = {\mathcal M}(G, \pi)$.

\noindent\textbf{end.}

\begin{lem}
\label{EdgePresentLemma}
Let $e = (u,v) \notin E(G)$. 
Let $G'$ be the output of \textbf{RAND}($G$). Then,
$$
\pp{e \in E(G')} = \frac{1}{2} \left( \frac{d(u)}{d(u) + 2} + \frac{d(v)}{d(v) + 2}\right) \le \frac{\Delta}{\Delta +2} .
$$
\end{lem}
\begin{pf}

  We have to estimate the probability that $u$ and $v$ are adjacent in $G'$.
  That is, $I(u) \cap I(v) \not= \emptyset$.

  Let $n_u \in N(u) \cup \{u\}$ be a vertex such that  $\pi(n_u) = \min_{w \in N(u) \cup \{u\}} \pi(w)$.
  Similarly,  let $n_v \in N(v) \cup \{v\}$ be a vertex such that  $\pi(n_v) = \min_{w \in N(v) \cup \{v\}} \pi(w)$. 

Clearly, $I(u) = [\pi(n_u), \pi(u)]$ and $I(v) = [\pi(n_v), \pi(v)]$. 
  Thus, it is easy to see that $I(u) \cap I(v) \not= \emptyset$ \emph{if}
  \textbf{(a)} $\pi(n_u) < \pi(v) < \pi(u)$ 
 \textbf{or}  \textbf{(b)} $\pi(n_v) < \pi(u) < \pi(v)$.
On the other hand,
 $I(u) \cap I(v) \not= \emptyset$ \emph{only if} either (a) or (b) hold.
To see this, first observe that $\pi(u) \not= \pi(w)$ for any $w \in N(v) \cup \{v\}$
and $\pi(v) \not= \pi(w)$ for any $w \in N(u) \cup \{u\}$. Without loss of generality,
let $\pi(u) < \pi(v)$. Now, it is obvious that $I(u) \cap I(v) \not= \emptyset$,
$\pi(n_v) < \pi(u) < \pi(v)$ since $\pi(n_v) \not= \pi(u)$.
Since (a) and (b) are mutually exclusive,
$$
\pp{e \in E(G')} = \pp{ \pi(n_u) < \pi(v) < \pi(u) } + \pp{\pi(n_v) < \pi(u) < \pi(v)}.
$$
We bound $\pp{\pi(n_u) < \pi(v) < \pi(u) }$ as follows.
  Let $X = \{u\} \cup N(u) \cup \{v\}$. Let $\pi_X$ be the projection of $\pi$ onto $X$.
Clearly, the event $\pi(n_u) < \pi(v) < \pi(u)$ translates to saying that
 $\pi_X(v) < \pi_X(u)$ and $\pi_X(v) \not= 1$. 
  Note that $\pi_X$ can be any permutation of $|X|$ elements with equal probability, which
 is $\frac{1}{(d(u) + 2)!}$. The number of permutations where $\pi_X(v) < \pi_X(u)$
equals $(d(u) + 2)!/2$. Moreover, the number of permutations where $\pi_X(v) = 1$ equals
$(d(u) + 1)!$. Note that the set of permutations with $\pi_X(v) = 1$ is 
a subset of the set of permutations with $\pi_X(v) < \pi_X(u)$.
It follows that
$$
\pp{\pi_X(v) < \pi_X(u)  \mbox{ and } \pi_X(v) \not= 1} =  \frac{(d(u) + 2) !/2 - (d(u) + 1)!}{(d(u) + 2)!} 
$$
which is $\frac{d(u)} { 2(d(u) + 2)}\enspace .$ Using similar arguments, it follows that \break
$\pp{\pi(n_v) < \pi(u) < \pi(v)} = \frac{d(v)}{2(d(v) + 2)}$. Summing the two bounds,
the result follows.
\end{pf}

\begin{lem}
\label{RandUpperLemma}
Let $I_1, I_2, \cdots, I_t$ be the output generated by $t$ invocations of
\textbf{RAND}($G$).
If $t \ge (\Delta + 2) \ln n$ then $\pp{E(G) =  E(I_1) \cap E(I_2) \cap \cdots
\cap E(I_t)} \ge \frac{1}{2}$.
\end{lem}
\begin{pf}
  For  $e = (u, v) \notin E(G)$,
  let $Z_e$ denote the event 
  $$
e \in E(I_1) \bigwedge e \in E(I_2)  \ldots \bigwedge e \in E(I_t)
$$ That is,
  $Z_e$ denote the event that $e \in E(I_1) \cap E(I_2) \cdots \cap E(I_t)$.
Note that for any $i, j \in \{1,\ldots,t\}$, $i \not= j$, the events $e \in E(I_i)$ and
$e \in E(I_j)$ are independent.   It follows from Lemma \ref{EdgePresentLemma} that
  $$
  \pp{Z_e} 
  ~= \prod_{j \in \{1,\ldots,t\}} \pp{e \in E(I_j)} 
  ~\le~ \left(\frac{\degr}{\degr+2}\right)^{t} \enspace.
  $$
  Hence by union bound,
  $$
  \pp{\bigvee_{e \notin E(G)} Z_e}
  \le~ \sum_{e\notin E(G)} \pp{Z_e}
  ~\le~ \frac{n^2}{2} \left(\frac{\degr}{\degr+2}\right)^{t} \,\le\, (n^2/2) e^{\frac{-2t}{\Delta + 2}} \enspace .
  $$
  If we choose $t = (\degr + 2)\ln n$ then the above probability is upper
  bounded by 1/2. 
\ignore{
If we choose $t = \frac{3}{2} (\degr + 2) \ln n$ the above probability is
upper bounded by $1/(2n)$.
}
Recall that $E(I_j) \subseteq E(G)$ for all $j$. It follows that $\pp{E(G) = \bigcap_{j} E(I_j)} = 1 - \pp{\bigvee_{e \notin E(G)} Z_e}$. Thus the result follows.
\end{pf}

\ignore{
As mentioned in the proof of Lemma \ref{RandUpperLemma}, if we fix $t = (\Delta + 2) \ln n$,
the resulting intersection graph is $G$ with probability at least $1/2$. Hence
}
Since the set of interval graphs generated by $\lceil(\Delta+2)\ln n\rceil$
invocations
of \textbf{RAND}$(G)$ is a valid interval graph representation of $G$ with
non-zero probability as shown above, we have the following corollary.

\begin{cor} \label{cor:det-upper}
  Let $G$ be a graph on $n$ vertices and with maximum degree $\Delta$. Then
  $
  \boxi(G) ~\le~ \lceil(\Delta + 2) \ln n\rceil \enspace .
  $ 
\end{cor}

\begin{lem}
\label{RandTimeLemma}
The \textbf{RAND} procedure can be implemented in $O(m + n)$ time 
assuming that a permutation of $\{1, \ldots, n\}$ can be generated uniformly at random in $O(n)$ time.
\end{lem}
\begin{pf}
The vertex set of $G$ is $\{1, \ldots, n\}$.
We make the standard assumption that $G$ is available as an adjacency list representation.
First generate $\pi$ and store the mapping $\pi^{-1}()$ in an array indexed from $1$ to $n$.
Maintain another array $D$ indexed from $1$ to $n$ such that $D[i] = 1$ if $l(i)$ for vertex
$i$ is already defined and $D[i] = 0$ otherwise.
We now construct $G' = {\mathcal M}(G, \pi)$ in $n$ steps as follows.
In $i$th step, consider the vertex $u = \pi^{-1}(i)$. 
Define $r(u) = i$.  If $D[u] = 0$ then $l(u) = i$.
For each element $w \in N(u)$, if $D[w] = 0$ then define $l(w) = i$ and assign $D[w] = 1$.
Thus total time taken in step $i$ is $O(|d(u)| + 1)$ steps. It follows that
the interval representation of $G'$ can be  constructed in  $O(m + n)$ steps.
\end{pf}

\begin{thm} \label{thm:det-upper}
  Given a graph $G$ on $n$ vertices and $m$ edges, with high probability, 
a box representation of
 $G$ in $\lceil(\Delta + 2) \ln n\rceil$ dimension can be constructed in 
$O(\Delta n^2  \ln^2 n )$ time, where $\Delta$ is the maximum degree of $G$.
\end{thm}
\begin{pf}
We construct an algorithm that takes the graph $G$ as input and tries to
compute an interval graph representation for $G$. It repeatedly
computes a set $S$ of $\lceil(\Delta+2)\ln n\rceil$ interval supergraphs 
of $G$ until it generates a set that is a valid interval graph representation
of $G$. If the algorithm fails to find such a set after $(\log_2 e)\ln n$
tries, it reports a failure. 
Each $S$ is computed through $\lceil(\Delta+2)\ln n\rceil$
invocations of \textbf{RAND}$(G)$. From lemma~\ref{RandUpperLemma}, the
probability of failure of this
algorithm is at most $(1/2)^{(\log_2 e)\ln n}=1/n$
and hence it computes a valid
interval graph representation of $G$ with high probability.
Using lemma~\ref{RandTimeLemma}, it is easily verified that
to compute a set of $(\Delta + 2) \ln n$ interval graphs, our algorithm
takes  $O(\Delta(m+n)\ln n)$ time. To verify whether a generated set of 
$(\Delta + 2) \ln n$ interval graphs is a valid interval representation or not,
it takes $O(n^2 \Delta \ln n)$ time. Thus the overall complexity is $O(\Delta n^2 \ln^2 n)$.
 
\end{pf}

\subsection{Almost tight example}\label{roberts}
We remark that for any given  $\Delta$  and $n > \Delta + 1$, we can construct
a graph $G$ on $n$ vertices and with maximum degree $\Delta$ such that
$\boxi(G) \ge \lfloor (\Delta + 2)/2 \rfloor$. We assume that $\Delta$ is even for the ease
of explanation. 
Roberts \cite{Roberts} has shown that for any even number $k$, there
exists a graph on $k$ vertices with degree $k-2$ and boxicity $k/2$.
We call such graphs as \emph{Roberts graphs}.
The Roberts graph on $n$ vertices is obtained by removing the edges 
of a perfect matching from a complete graph on $n$ vertices.  We take
such a graph by fixing $k = \Delta + 2$ and we let the remaining $n- (\Delta +2)$ vertices
be isolated vertices. Clearly, the boxicity of such a graph is also $k/2 = (\Delta + 2)/2$,
whereas the maximum degree is $\Delta$. Thus our upper bound is tight up to a factor
of $2 \ln n$.

\section {Derandomization}

 In this section we derandomize the above randomized algorithm to
 obtain a deterministic polynomial time algorithm to output the 
 box representation in $(\Delta+2)  \ln n$ dimensional space
 for a given graph $G$ on $n$ vertices with maximum degree $\Delta$.
 
 \begin {lem}
 \label {DerandomizationBasicLemma}
 Let $G=(V, E)$ be the graph. Let $E(\overline G)$
 be the edge set of the complement of $G$. Let $H \subseteq E(\overline G)$. 
 Then we can construct an interval super graph $G'$ of $G$ in polynomial time 
such that
 $|E(\overline {G'}) \cap H | \ge \frac {2}{\Delta +2} |H|$. 
 \end {lem}
\begin {pf}
We derandomize the \textbf{RAND} algorithm
to devise a deterministic algorithm to construct $G'$.




Our deterministic strategy defines a permutation $\pi$ on the vertices $\{1, \ldots, n\}$
of $G$. The desired $G'$ is then obtained as ${\mathcal M}(G, \pi)$.
Let the ordered set $V_n = < v_1, \ldots, v_n>$ denote the final permutation given by $\pi$.
We construct $V_n$ in a step by step fashion. At the end of step $i$,
we have already defined the first $i$ elements of the permutation,
namely the ordered set $V_i  = <v_1, \ldots, v_i>$, where each $v_j$ is distinct.
Let $V_0$ denote the empty set.
Having obtained $V_i$ for $i\ge 0$, we compute $V_{i+1}$ in the next step as follows.

 Given an ordered set $V_i$  of $i$  vertices 
 $<v_1,v_2,\ldots,v_i>$,  let $V_i \diamond u$ denote the ordered 
set of the $i+1$ vertices   $<v_1,v_2,\ldots,v_i,u>$. (We will abuse
notation and use $V_i$ to denote the underlying unordered set also,
when there is no chance of confusion.) Let $V_0 \diamond u$ denote $<u>$.

Consider an invocation of  the randomized algorithm \textbf{RAND} whose output is 
denoted as $G''$.
For each $e \in H$, let $x_e$ denote the indicator random variable which is $1$
if $e \notin E(G'')$, and $0$ otherwise. Let $X_H = 
\sum_{e \in H} x_e$.

Given an ordered set $S = <u_1,\ldots,u_r>$, let
$\mathcal{Z}(S)$  denote the event that the first $r$ elements
of the random permutation generated by \textbf{RAND} is given
by the ordered set $S = <u_1, \ldots, u_r>$. 
Note that $\pp{\mathcal{Z}(V_0)} = 1$ since the first $0$ elements of any permutation
is the empty set $V_0$.

Let  $x_e|\mathcal{Z}(V_i)$ denote the indicator random variable corresponding
to $x_e$
conditioned on the event $\mathcal{Z}(V_i)$.

Similarly, let the random variable $X_H|\mathcal{Z}(V_i)$ denote $|E(\overline{G''}) \cap H|$
conditioned on the event $\mathcal{Z}(V_i)$.

\noindent
For $i \ge 0$,
let $f_e(V_i)$ denote $\pp{x_e = 1 ~|~ \mathcal{Z}(V_i)}$
and let $F(V_i)$  denote $\pe{X_H ~|~ \mathcal{Z}(V_i)}$.

Note that $f_e(V_0)$ denote $\pp{x_e = 1}$
and $F(V_0)$ denote $\pe{X_H}$.

Clearly 
$$
F(V_i) = \sum_{e \in H} f_e(V_i).
$$

By Lemma \ref{EdgePresentLemma}, we know that for any $e \in H$, $f_e(V_0) \ge \frac{2}{\Delta + 2}$.
Thus $F(V_0) \ge \frac{2|H|} {\Delta + 2} \enspace.$
Clearly, 
$$
\pe {X_H | \mathcal{Z}(V_i)} = \frac {1}{|V - V_i|} \sum_{u \in V - V_i} \pe {X_H|\mathcal{Z}(V_i \diamond u)}.
$$ 

Let $u \in V- V_i$ be such that 
$$
\pe{X_H  | \mathcal{Z}(V_i \diamond u)} = \max_{w \in V - V_i} \pe{X_H | \mathcal{Z}( V_i \diamond w) }.
$$
Define $V_{i+1} = V_i \diamond u$. It follows that
 $$
  F(V_{i+1}) = \pe{X_H|\mathcal{Z}(V_{i+1})} \ge \pe{X_H | \mathcal{Z}(V_i) } = F(V_i).
$$
In particular, it is also true that $F(V_1) \ge F(V_0)$.

After $n$ steps,
we obtain the final permutation $V_n$. 
Applying the above inequality $n$ times,
it follows that
$$
\pe{X_H | \mathcal{Z}(V_n)} = F(V_n) \ge F(V_0) .
$$
Recalling that $F(V_0) \ge \frac{2|H|}{\Delta + 2}$, we have $F(V_n) \ge \frac{2|H|}{\Delta + 2}$.

Let $\pi$ be the permutation that corresponds to the ordered set $V_n$.
The final interval super graph $G'$ output by our deterministic strategy is ${\mathcal M}(G, \pi)$.
Note that, $F(V_n) =  \pe{X_H|\mathcal{Z}(V_n)}$ is the same as $|E(\overline{G'}) \cap H|$.
Thus we have shown that $|E(\overline{G'}) \cap H| \ge \frac{2}{\Delta + 2}|H|$ as claimed. 

It remains to show that the above deterministic strategy takes only polynomial time.
This can be shown as follows using  Lemma \ref{CondProbEstimationLemma} (see below). 
Recall that, given a vertex $w \in V - V_i$, $F(V_i \diamond w)$ is simply $\sum_{e \in H} f_e(V_i \diamond w)$.
It follows from Lemma \ref{CondProbEstimationLemma} that $F(V_i \diamond w)$ can be
computed in polynomial time.
Recall that given $V_i$, $V_{i+1}$ is $V_i \diamond u$ where $u$ maximizes 
$F(V_i \diamond w)$ where $w \in V - V_i \enspace.$ Clearly such a $u$
can also be found in polynomial time. Since there are only $n$ steps before computing $V_n$,
the overall running time is still polynomial.
\end {pf}

\begin{lem}
\label{CondProbEstimationLemma}
For any ordered set $V_i = <v_1, \cdots, v_i>$ and any $e \in H$,  $f_e(V_i)$ 
can be computed  in polynomial time.
\end{lem}

\begin{pf}
 Let $e = (u,v)$. We can  compute $f_e(V_i)$ as follows:

 The easiest case is when both $u,v$ are  in $V_i$. In this case, the intervals corresponding
 to $u$ and $v$, namely $I(u)$ and $I(v)$ are already defined. (Recall how ${\mathcal M}(G,\pi)$ defines an interval supergraph of $G$. See definition \ref{mmap}). Therefore either $e \in E(G')$
 or $e \notin E(G')$. Therefore the conditional probability is either 0 or 1. This is summarized as:

\begin {center}
 \noindent
 \emph{Case 1.1:   $u,v \in V_i$ and  $I(u) \cap I(v) = \emptyset$. }

Then $f_e(V_i) = 1$ since $(u, v) \notin E(G')$.

 \noindent
 \emph{Case 1.2:   $u,v \in V_i$ and $I(u) \cap I(v) \ne \emptyset$: }

Then $f_e(V_i) = 0$ since $(u, v) \in E(G')$.
\end {center}

Next let us consider the case $u \in V_i$ and $v \notin V_i$. In this case 
 $I(u)= [l(u),r(u)]$ is already defined but $I(v)$ is not yet determined since $r(v)$ is 
still random.
But it is clear that $r(v) > r(u)$. Moreover, we can 
determine whether  $l(v)< r(u)$ or not, irrespective of $r(v)$:  If
there is a vertex $w \in N(v) \cap V_i$ such that $r(w) < r(u)$, then $l(v) < r(u)$,
otherwise $l(v) > r(u)$.  Thus in this case also the conditional probability is either $0$ or $1$,
as computed by the following procedure:

\begin {center}
 \emph{Case 2.1: $u \in V_i, v \notin V_i$ and $N(v) \cap V_i = \emptyset$:}

 Then $f_e(V_i) = 1$.

 \emph{Case 2.2: $u \in V_i, v \notin V_i$ and $N(v) \cap V_i \ne \emptyset$: }

Let $t = \min_{w \in N(v) \cap V_i} r(w) \enspace .$\\
\smallskip
     If $r(u) > t$ then $f_e(V_i) = 0$  else $f_e(V_i) = 1$.
\end{center}

 Now we examine the case when both $u$ and $v$ do not belong to $V_i$.
First consider the sub case when both $N(u) \cap V_i$ and $N(v) \cap V_i$  are non-empty.
Then clearly $I(u) \cap I(v) \not= \emptyset$. This is because of the following.
Let us denote by $v_i$, the $i$th vertex (the last vertex) in $V_i$. Clearly,
$l(u) \le r(v_i) < r(u)$ and $l(v) \le r(v_i) < r(v)$. Thus $r(v_i) \in I(u) \cap I(v)$.
It follows that in this case the conditional probability is $0$ as give below.

\begin{center}
\emph{Case 3.1:  $u, v \in V - V_i$ and $N(v) \cap V_i \ne \emptyset$ and  $N(u) \cap V_i \ne \emptyset$: }

      Then $f_e(V_i) = 0$. 
\end{center}

Next subcase is when $N(u) \cap V_i$ and $N(v) \cap V_i$ are both empty.
That is, the set $X = \{u\} \cup N(u) \cup \{v\} \cup N(v)$ has empty
intersection with $V_i$.  
Let $\pi_X$ be the projection of $\pi$ onto $X$.
Since $X \cap V_i = \emptyset$, $\pi_X$ can be any possible
permutation of $|X|$ elements with equal probability, namely $\frac{1}{|X|!}$.
We estimate the conditional probability $f_e(V_i)$ as follows.
Clearly, $(u, v) \notin E(G')$ if and only if $\pi_X(u) < \min_{w \in \{v\} \cup N(v)} \pi_X(w)$
or $\pi_X(v) < \min_{w \in \{u\} \cup N(u)} \pi_X(w)$.
It is easy to see that the probability for the above condition to hold
is
$ \frac{1}{d(u) + 2} + \frac{1}{d(v) + 2} \enspace \cdot$
Thus, we have the following case:

\begin{center}
\emph{Case 3.2:  $u,v \in V - V_i$ and $N(v) \cap V_i = \emptyset$ and $N(u) \cap V_i = \emptyset$: }

Then
$ f_e(V_i) =  \frac{1}{d(u) + 2} + \frac{1}{d(v) + 2} \enspace \cdot $
\end{center}

Now we are left with the last sub case:  $N(v) \cap V_i \not= \emptyset$ and
$N(u) \cap V_i = \emptyset$. Let $X = \{u\} \cup N(u) \cup \{v\}$.
Note that $X \cap V_i = \emptyset$. Clearly 
$\pi_X$ can be any possible permutation of $|X|$ elements with equal probability, 
namely $\frac{1}{|X|!} = \frac{1}{(d(u) + 2)!}\enspace .$
We estimate the conditional probability $f_e(V_i)$ as follows.

We claim that that $(u, v) \notin E(G')$
if and only if $\pi_X(v) = 1$. To see this,
first observe that if $(u, v) \notin E(G')$ then $\pi_X(v) < \pi_X(u)$ because
 otherwise $r(u) < r(v)$ and since $N(v) \cap V_i \not= \emptyset$, we get further 
$l(v) < r(u) < r(v)$ leading to a contradiction. 
Therefore $1 \le \pi_X(v) < \pi_X(u)$.
Now, if $\pi_X(w) = 1$ for some $w \in N(u)$ then
$l(u) < r(v) < r(u)$ and thus $I(u) \cap I(v) \not= \emptyset$ leading to a contradiction.
On the other hand, if $\pi_X(v) = 1$ then clearly,
$r(v) < r(w)$ for any $w \in N(u) \cup \{u\}$. It follows that
$r(v) < l(u)$ and therefore $I(v) \cap I(u) = \emptyset$.

Now, having shown that $(u, v) \notin E(G')$ if and only if $\pi_X(v) = 1$, we can compute
$f_e(V_i)$ by computing the conditional probability that $\pi_X(v) = 1$, which is
simply $\frac{(d(u) + 1)!}{(d(u) + 2)!} = \frac{1}{d(u) + 2}$.
Thus we have the following final case:

\begin{center}
\emph{Case 3.3:  $u,v \in V - V_i$ and $N(v) \cap V_i \not= \emptyset$ and $N(u) \cap V_i = \emptyset$: }

Then
$ f_e(V_i) = \frac{1}{d(u) + 2} $.
\end{center}

\end{pf}

 \begin {thm}
  \label {DerandFinalTheorem}
Let $G$ be a graph on $n$ vertices with maximum degree $\Delta$.
 The box representation of $G$ in 
$ \lceil (\Delta + 2) \ln n \rceil$ dimension can be constructed in polynomial time.
\end {thm} 

\begin {pf}
 Let $h = |E(\overline{G})|$.
  It follows from Lemma \ref{DerandomizationBasicLemma} that
   we can construct $t$ interval graphs such that the number of edges of 
  $E(\overline{G})$ which is present  in all of these $t$ interval graphs is
  at most $\left ( \frac {\Delta}{\Delta +2} \right )^t h$. 
If  $\left ( \frac {\Delta}{\Delta +2} \right )^t h < 1$, then we are done.  That is, 
 we are done if $t \ln \left ( \frac {\Delta}{\Delta + 2} \right ) + \ln h < 0$ is
 true. Clearly this is true, if $t >  {\ln h}/{\ln \left ( \frac {\Delta +2}
  {\Delta} \right ) } $. We can assume that $\Delta > 1$.
If $\Delta \le 1$ then $G$ is an interval graph and the theorem trivially holds.
Using the fact that $\ln \frac {\Delta +2 }{\Delta} > 
  \frac {2}{\Delta} - \frac {1}{2} (\frac {2}{\Delta} )^2$, we obtain 
   $\boxi (G) \le  \frac { \Delta^2}{2 (\Delta - 1)} \ln h \le (\Delta + 2) 
  \ln n$. By Lemma \ref{DerandomizationBasicLemma},
each interval graph is constructed in polynomial time.
  Hence the total running time is still polynomial. Thus the theorem follows.
\end {pf}

\section{In terms of average degree}

It is natural to ask whether our upper bound of $(\Delta + 2) \ln n$ still
holds, if we replace $\Delta$ by the average degree $d_{av}$.
Unfortunately this is not true in general: we show below an infinite family of graphs
where the boxicity is exponentially higher than $(d_{av} + 2) \ln n$.
Construct a  graph $G$ on $n$ vertices as follows. First take a Roberts graph
on $n_1$ vertices where $n_1 \le n$. (Refer to section \ref{roberts} for the definition
of Roberts graph.) Let the remaining $n - n_1$ vertices form a path which is connected by an edge to 
one of the vertices of the Roberts graph.
The average degree of $G$ is $d_{av}  = (n_1 ( n_1 - 2) + 2(n - n_1))/n$, whereas its
boxicity is at least $n_1/2 \ge \frac{1}{2}\sqrt{n (d_{av} - 2)} \enspace .$
It is easy to see that the boxicity of $G$ is exponentially larger than $(d_{av} + 2) \ln n$
for example when $n_1 = \theta(\sqrt{n})$.

Nevertheless we show the following general upper bound for boxicity in terms of $d_{av}$.
In fact, we will express the upper bound in terms of $n$ and the number of edges $m$.

\begin{thm}
For a connected graph $G$ on  $n$ vertices and $m$ edges,  $\boxi(G) \le  5\sqrt{m \ln n}$.
Moreover, there exists a connected graph $G$ with $n$ vertices and $m$ edges such that
$\boxi(G) ~\ge   \sqrt {\frac {m -n} {2}}$. 
\end{thm}

\begin{pf}
We show the upper bound as follows.
Let $x = \sqrt{m/\ln n}$. Let $V'$ denote the set of vertices in $G$
whose degree is at least $x$. It is straightforward to verify that
$|V'| \le {2m }/{x}$.
Let $G''$ be the induced subgraph  on $V - V'$.
Each vertex in $G''$ has degree at most $x$. 
By Corollary \ref{cor:det-upper}, we obtain that $\boxi(G'') \le (x + 2) \ln n$.
Since  $\boxi(G'') + |V'|$
is a trivial upper bound for $\boxi(G)$,
it follows that $\boxi(G) \le  (x +2) \ln n + 2m/{x} \le 5 \sqrt{m \ln n}$ since $m \ge n-1$.
The example graph discussed in the beginning of this section serves as the example that
illustrates the lower bound.
\end{pf}

\subsection{Boxicity of Random Graphs}

Though in general boxicity of a graph is not upper bound by $(d_{av} + 2)
\ln n$, where $d_{av}$ is its average degree,
we now show that for almost all graphs, the boxicity is $O(d_{av}\ln n)$.

We show that for almost all graphs in the
the $\mathcal{G}(n,m)$ model, $\boxi(G)$ is $O(c\ln n)$ where $c=2m/n$.
We shall only consider connected graphs here and so assume that $m\geq n-1$.
Thus, we have, $c>1$.
But we first show the result for the $\mathcal{G}(n,p)$ model setting
$p=c/(n-1)$. As shown in
\cite{Bol01}, we can then carry over the result to the $\mathcal{G}(n,m)$
model since $p=m/\binom{n}{2}$.

Consider the $\mathcal{G}(n,p)$ model with $p=c/(n-1)$.
Let $G$ denote a random graph drawn \com{on} according to this model.
For a vertex $u$, define a random variable $d_u$ that denotes the degree
of $u$, i.e., $d_u=|N(u)| \com{$.  $d_u} = \sum_{v\in V(G), v\not= u} e_{u,v}$ where
$e_{u,v}$ is an indicator
random variable whose value is 1 if $(u,v)\in E(G)$ and 0 otherwise.
Therefore, $\pe{d_u}=p(n-1)=c$.

\noindent\textit{Case 1:} $c\geq \ln n$.

Since $d_u$ is the sum of independent Bernoulli random variables, we can
use Chernoff bound to bound the probability of $d_u$ becoming large. In
particular, we use the following form of the Chernoff bound given in
\cite{AV79} \com{added this} for the rest of the proof.
\begin{eqnarray}\label{cher}
\pp{X\geq (1+\delta)\pe{X}} ~~\leq~~ e^{-\frac{\delta^2 \pe{X}}{2+\delta}}
\end{eqnarray}
for all $\delta>1$. Taking $\delta=5$, we get,
$\pp{d_u\geq 6c}\leq {1}/{n^3}$.
\com{
Now, we can show the following bound on $\Delta(G)$.
$$\pp{\Delta(G)\geq 6c}=\pp{\exists u\in V(G), d_u\geq 6c}
\leq\sum_{u \in V(G)}\pp{d_u\geq 6c}= {1}/{n^2}$$}
Now, by union bound, it follows that
$\pp{\Delta(G)\geq 6c}=\pp{\exists u\in V(G), d_u\geq 6c}
\leq {1}/{n^2}.$
Using the result $\boxi(G)\leq(\Delta+2)\ln n$, we now have,
$\boxi(G)\leq (6c+2)\ln n$ with probability at least $1-{1}/{n^2}$.

\noindent\textit{Case 2:} $c<\ln n$.

Let $S_u=V(G)-N(u)-\{u\}$.\\
Let $N'(u)=\{v\in S_u ~|~ \exists u'\in N(u)\mbox{ such that }(u',v)\in
E(G)\}$.\\
In this case, we will use a different technique to upper bound boxicity.
\com{added some explanation for square of G}
Let the graph $G^2$ denote the square of $G$. That is, $V(G^2) = V(G)$ and $(u, v) \in E(G^2)$
if there is a path of length $1$ or $2$ between $u$ and $v$.
\com{It is shown in} The authors showed in \cite{CFNMaxdeg} that the boxicity of a graph $G$ is at most
$2\Delta(G^2)+2$. We will show below that
if $c<\ln n$, then $\Delta(G^2)\leq c+6\ln n+7c^2+42c\ln n$, with high
probability. The reader may note that the degree of a vertex $u$ in $G^2$
equals $|N(u)|+|N'(u)|$. 
\com{Added the following to give the direction..}
We will now show that for any vertex $u$,
$\pp{|N(u)| + |N'(u)| \notin O(c \log n)} \le 3/n^3$.

Let $k=c+6\ln n$. 
We apply Chernoff bound (\ref{cher}) with $\delta = {6 \ln
n}/{c}$
to obtain
\com{Fix $\delta=\frac{6\ln n}{c}$. (Thus we have $\delta\geq 6$).}
\begin{eqnarray*}\label{eq1}
\pp{d_u\geq k} ~~\leq~~ e^{-{\delta(6\ln n)}/{(2+\delta)}}
~~\leq~~ {1}/{n^3}
\end{eqnarray*}
\com{Consider the case when $d_u=|N(u)|<k$.}
\com{For some subset $A$ of $V(G)$ such that $|A|<k$, let $Z(A)$ denote the event
that $N(u)=A$.
Assume that $Z(A)$ has happened, i.e., $N(u)=A$.}
Let $A \subseteq V(G)$ such that $|A| < k$. Let $Z(A)$
denote the event that $N(u) = A$.
Now, for each vertex $v\in S_u$,
let $X_{v,A}$ denote an indicator random variable indicating whether $v\in
N'(u)$ conditioned on the event $Z(A)$. 
\com{Thus,\\
$X_{v,A}=\left\{\begin{array}{l}
1, \mbox{ if }v\in N'(u)\mbox{, provided }Z(A)\mbox{ occurred.}\\
0, \mbox{ otherwise.}\\
\end{array}\right.$\\}
Note that for any vertex $v\in S_u$, $\pp{X_{v,A}=1}\leq kp$.
Let $X_A=\sum_{v\in S_u} X_{v,A}$. 
It follows that $\pe{X_A}\leq kp(n-1)=kc$. Since $X_A$ is the sum of
independent
Bernoulli random variables, we apply the Chernoff bound
(\ref{cher}) by
fixing  $\delta={6kc}/{\pe{X_A}}$ to obtain
$\pp{X_A\geq 7kc} ~~\leq~~ e^{-{\delta(6kc)}/{(2+\delta)}}
~~\leq~~ {1}/{n^3}$.

\com{
	again. Thus,\\
	$$\pp{X_A\geq (1+\delta)\pe{X_A}}\leq e^{-\frac{\delta^2 \pe{X_A}}
	{2+\delta}}$$
	Put $\delta=\frac{6kc}{\pe{X_A}}$ (thus, $\delta\geq 6$).
	\begin{eqnarray*}
	\pp{X_A\geq 7kc}&\leq&\pp{X_A\geq\pe{X_A}+6kc}\\
	&&(\mbox{since }\pe{X_A}\leq kc)\\
	&\leq&e^{-\frac{\delta^2\pe{X_A}}{2+\delta}}=e^{-\frac{\delta(6kc)}
	{2+\delta}}\\
	&\leq& e^{-3kc}~~~(\mbox{substituting }2\delta\mbox{ for }
	2+\delta)\\
	&\leq&e^{-3\ln n}~~~(\mbox{since }kc=c^2+6c\ln n\geq\ln n)\\
	&\leq&\frac{1}{n^3}
	\end{eqnarray*}
}

Let the random variable $X_u = |N'(u)|$.
\com{ denote the random variable that takes the value $|N'(u)|$.}
We now have,
\begin{eqnarray*}
\pp{X_u\geq 7kc~|~d_u<k}&=&\sum_{A\subseteq V(G),|A|<k}
\pp{(X_u\geq 7kc)\wedge Z(A)}\\
&=&\sum_{A\subseteq V(G),|A|<k}\pp{X_A\geq 7kc}\pp{Z(A)} ~\le~ 1/n^3
\end{eqnarray*}
It follows that
\begin{eqnarray*}
\pp{X_u\geq 7kc}&=&\pp{X_u\geq 7kc~|~d_u<k}\pp{d_u<k}\\
& &+\pp{X_u\geq 7kc~|~d_u\geq k}\pp{d_u\geq k}\\
&\leq&({1}/{n^3})\pp{d_u<k} + ({1}/{n^3})\pp{X_u\geq
7kc~|~d_u\geq k} ~\le~ 2/n^3
\end{eqnarray*}
Let $t_u=|N(u)|+|N'(u)|=d_u+X_u$. Combining the bounds on the
values of $d_u$ and $X_u$, we get,
$$\pp{t_u\geq k+7kc} ~\leq~ \pp{d_u\geq k}+\pp{X_u\geq 7kc}
~\leq~ {3}/{n^3}$$
Observe that $\Delta(G^2) = \max_{u \in G} t_u$.
Thus, by
applying union
bound, we obtain
\begin{eqnarray*}
\pp{\Delta(G^2)\geq k+7kc}&=&\pp{\bigvee_{u\in V(G)} t_u
\geq k+7kc} ~\leq~ {3}/{n^2}
\end{eqnarray*}
Thus, with high probability,
$\Delta(G^2)<k+7kc=c+6\ln n+7c^2+42c\ln n$.
Recalling that $\boxi(G)\leq 2\Delta(G^2)+2$, we obtain
$\boxi(G) \in O(c \ln n)$ with high probability, since $c<\ln n$.
\com{ The almost all graphs statement cannot be made in the
G(n,p) model ??
$\leq 2c+12\ln n+14c^2+84c\ln n$ with high probability. Since $c<\ln n$,
this implies that $\boxi(G)$ is $O(c\ln n)$ for almost all
$G$. }

Having shown that in the $\mathcal{G}(n,p)$ model,
$\pp{\boxi(G)\not\in O(c\ln n)}\leq {3}/{n^2}$,
the following relation from page 35 of \cite{Bol01} helps us to extend our
result to the $\mathcal{G}(n,m)$ model.
$$P_m(Q)\leq 3m^{1/2}P_p(Q)$$
where $Q$ is a property of graphs of order $n$, and $P_m(Q)$ and $P_p(Q)$ are
the probabilities of a graph chosen at random from the $\mathcal{G}(n,m)$
or the $\mathcal{G}(n,p)$ models respectively to have property $Q$ given that
$p=m/\binom{n}{2}$.
Using this result, we now have, for a graph $G$ drawn randomly from the
$\mathcal{G}(n,m)$ model,
$$\pp{\boxi(G)\not\in O(c\ln n)}\leq 9 n^{-2} \sqrt{m}\leq
{9}/{n}$$
As $c=2m/n = d_{av}$, which is the average degree, 
we have shown that for almost all graphs
with a given average degree $d_{av}$, the boxicity  is $O(
d_{av} \ln n)$.

Thus we have the following theorem:

\begin{thm}
\label{RandGraphThm}
  For a random graph $G$ on $n$ vertices and $m$ edges drawn according to $\mathcal{G}(n,m)$
model,
  $$
    \pp{\boxi(G)  ~=~ O \left(\frac {2m}{n} \ln n \right) } \ge 1 - \frac{9}{n}  \enspace
  $$
\end{thm}

\section {Acknowledgement}

We thank the anonymous referee for carefully reading the paper and pointing out 
a serious mistake in the last section  of the first version of the paper.


\begin{thebibliography}{23}

\bibitem{Chan03}
{\sc T.~M. Chan}, {\em Polynomial-time approximation schemes for packing and
  piercing fat objects}, Journal of Algorithms, 46 (2003), pp.~178--189.

\bibitem{Agarwal98}
{\sc P.~K. Agarwal, M.~van Kreveld, and S.~Suri}, {\em Label placement by
  maximum independent set in rectangles}, Comput. Geom. Theory Appl., 11
  (1998), pp.~209--218.

\bibitem{Bellantoni}
{\sc S.~Bellantoni, I.~B.-A. Hartman, T.~Przytycka, and S.~Whitesides}, {\em
  Grid intersection graphs and boxicity}, Discrete mathematics, 114 (1993),
  pp.~41--49.

\bibitem{Berman2001}
{\sc P.~Berman, B.~DasGupta, S.~Muthukrishnan, and S.~Ramaswami}, {\em
  Efficient approximation algorithms for tiling and packing problems with
  rectangles}, J. Algorithms, 41 (2001), pp.~443--470.

\bibitem{Bodland3}
{\sc H.~L. Bodlaender}, {\em A tourist guide through treewidth}, Acta
  Cybernetica, 11 (1993), pp.~1--21.

\bibitem{Bol01}
{\sc B.~Bollob\'as}, {\em Random Graphs}, Cambridge University Press, 2~ed.,
  2001.

\bibitem{CN05}
{\sc L.~S. Chandran and N.~Sivadasan}, {\em Treewidth and boxicity}.
\newblock Submitted, Available at http://arxiv.org/abs/math.CO/0505544.

\bibitem{ChangWest}
{\sc Y.~W. Chang and D.~B. West}, {\em Rectangle number for hyper cubes and
  complete multipartite graphs}, in 29th {SE} conf. Comb., Graph Th. and Comp.,
  Congr. Numer. 132(1998), 19--28.

\bibitem{ChangWest1}
\leavevmode\vrule height 2pt depth -1.6pt width 23pt, {\em Interval number and
  boxicity of digraphs}, in Proceedings of the 8th International Graph Theory
  Conf., 1998.

\bibitem{Coz}
{\sc M.~B. Cozzens}, {\em Higher and multidimensional analogues of interval
  graphs}.
\newblock Ph. D thesis, Rutgers University, New Brunswick, NJ, 1981.

\bibitem{Erl01}
{\sc T.~Erlebach, K.~Jansen, and E.~Seidel}, {\em Polynomial-time approximation
  schemes for geometric intersection graphs}.
\newblock To appear in SIAM Journal of Computing.

\bibitem{Feinberg}
{\sc R.~B. Feinberg}, {\em The circular dimension of a graph}, Discrete
  mathematics, 25 (1979), pp.~27--31.

\bibitem{Has99}
{\sc J.~Hastad}, {\em Clique is hard to approximate within $n^{1-\epsilon}$},
  Acta Mathematica, 182 (1998), pp.~105--142.

\bibitem{Klok2}
{\sc T.~Kloks}, {\em Treewidth: Computations And Approximations}, vol.~842 of
  Lecture Notes In Computer Science, Springer Verlag, Berlin, 1994.

\bibitem{Kratochvil}
{\sc J.~Kratochvil}, {\em A special planar satisfiability problem and a
  consequence of its {NP}--completeness}, Discrete Applied Mathematics, 52
  (1994), pp.~233--252.

\bibitem{Roberts}
{\sc F.~S. Roberts}, {\em Recent Progresses in Combinatorics}, Academic Press,
  New York, 1969, ch.~On the boxicity and Cubicity of a graph, pp.~301--310.

\bibitem{Scheiner}
{\sc E.~R. Scheinerman}, {\em Intersectin classes and multiple intersection
  parameters}.
\newblock Ph. D thesis, Princeton University, 1984.

\bibitem{Shearer}
{\sc J.~B. Shearer}, {\em A note on circular dimension}, Discrete mathematics,
  29 (1980), pp.~103--103.

\bibitem{Thoma1}
{\sc C.~Thomassen}, {\em Interval representations of planar graphs}, Journal of
  combinatorial theory, Ser {B}, 40 (1986), pp.~9--20.

\bibitem{TroWest}
{\sc W.~T. Trotter and J.~D.~B. West}, {\em Poset boxicity of graphs}, Discrete
  Mathematics, 64 (1987), pp.~105--107.

\bibitem{Yan1}
{\sc M.~Yannakakis}, {\em The complexity of the partial order dimension
  problem}, {SIAM} Journal on Algebraic Discrete Methods, 3 (1982),
  pp.~351--358.

\bibitem{AV79}
{\sc D.~Angluin and L.~G. Valiant}, {\em Fast probabilistic algorithms for
  hamiltonian circuits and matchings.}, J. Comput. Syst. Sci., 18 (1979),
  pp.~155--193.

\bibitem{CFNMaxdeg}
{\sc L.~Sunil Chandran, Mathew C.~Francis, and Naveen Sivadasan},
{\em Boxicity and maximum degree}.
\newblock To appear in Journal of Combinatorial Theory, Ser. B.

\end{thebibliography}
\end{document}